\documentclass[twocolumn,pra,showpacs,floatfix,groupedaddress]{revtex4}

\usepackage{amsmath,amssymb,amsfonts,bm}
\usepackage[usenames,dvipsnames]{color}
\usepackage{graphicx}
\usepackage{multirow}
\usepackage{comment}

\usepackage{color}

\definecolor{Remarks}{rgb}{1,0.3,0.3}
\newcommand\COMMENTED[1] {}

\newcommand{\Deltam}{\bm{\Delta}}
\newcommand{\Hm}{\mathbf{H}}
\newcommand{\cm}{\mathbf{c}}
\newcommand{\qm}{\mathbf{q}}
\newcommand{\rb}{\mathbf{r}}

\begin{document}

\title{FFLO order in ultra-cold atoms in three-dimensional optical lattices}

\author{Peter Rosenberg}

\author{Simone Chiesa}

\author{Shiwei Zhang}
\affiliation{Department of Physics, College of William \& Mary,
  VA 23188, USA} 

\begin{abstract}
We investigate different ground-state phases of attractive spin-imbalanced populations of 
fermions in 3-dimensional optical lattices. 
Detailed numerical calculations are performed
using Hartree-Fock-Bogoliubov theory 
to determine the ground-state properties systematically for different values of density, spin polarization 
and interaction strength. We first consider the high density and low polarization regime, in which
the effect of the optical lattice is most evident. We then proceed to the low density and high polarization 
regime where the effects of the underlying lattice are less significant and the system begins to resemble 
a continuum Fermi gas. We explore the effects of density, polarization and interaction on the character
of the phases in each regime and highlight the qualitative differences between the two regimes.
In the high-density regime, the order is found to be of Larkin-Ovchinnikov type, linearly oriented 
with one characteristic wave vector but varying 
in its direction with the parameters. 
At lower densities the order parameter develops more structures involving multiple wave vectors. 
\end{abstract}

\maketitle

\section{Introduction}

In the past two decades remarkable progress in cold atom physics has
opened a new frontier in the construction and precise control of quantum
systems. Following the development of a number of important experimental 
techniques, including Feshbach resonances and optical lattices, it was quickly 
suggested that ultra-cold atomic gases provide an ideal setting for the realization 
and investigation of a variety of exotic physical phenomena \cite{Hofstetter2002}. 
These systems 
provide experimental analogues to many condensed matter systems,
but are also highly tunable and free of disorder. These experiments represent an 
exciting opportunity to simulate the fundamental mechanisms and models of
condensed matter physics, for instance Cooper pairing of fermions and
the Hubbard model, without the additional complexities presented by real 
materials. A number of experiments have already demonstrated the possibilities 
for ultra-cold atomic gases, including inducing superfluidity in fermionic systems 
and probing the BEC-BCS crossover \cite{Regal2004,Shin2006, Kohl2005, Stoferle2006}.

In light of these advances, one system that has attracted considerable interest is an 
ultra-cold atomic gas
in an optical lattice with unequal populations of two hyperfine 
states. The hyperfine states can be seen as two distinct spin species, and an attractive 
interaction can be induced between them, with its strength tunable, using a Feshbach resonance. This system 
represents an experimental simulation of the attractive fermionic Hubbard model. It was 
first suggested by Fulde and Ferrell (FF)\cite{Fulde1964}, and separately by Larkin and 
Ovchinnikov (LO)\cite{Larkin1965}, that the mismatched Fermi surfaces in a polarized system 
of this type could result in an instability to the formation of a condensate of finite-momentum 
electron pairs. However, the FFLO phase has eluded conclusive detection for nearly fifty years. 
Considering how challenging the observation of this phase has proven to be, reliable determination
of the parameter domain in which this phase might exist, and its properties, remains an 
important goal.

Many efforts have been made, using a variety of theoretical and numerical techniques, to achieve 
this goal and to characterize the properties of the FFLO phase.  However, in most cases these studies 
were limited to targeted states, fixed size simulation cells or to one- and two-dimensional lattices \cite{
Koponen2006, Koponen2007, 2Dattractive,Chen2009,Loh2010}. 
Three-dimensional lattices are in many ways the most direct and natural for optical lattice experiments with ultra-cold 
atomic gases,
so these systems offer the most realistic 
possibility of observing FFLO states. With this in mind, we
map the density-polarization 
phase diagram for spin-imbalanced fermions with attractive interactions
in a 3D optical lattice in the present study.

While 3D systems may present great opportunities to observe the FFLO state experimentally,
they present a considerable computational challenge. We carry out detailed calculations using the Hartree-Fock-Bogoliubov theory,
which is the simplest quantitative approach. At the minimum, results from these mean-field calculations 
provide a qualitative understanding of the nature of the phases in a large region of the parameter space, and
propose candidate phases for more 
elaborate (and computationally intensive) many-body approaches. In fact, experience \cite{HubHF2D-Xu-2011,SDW-Hub2D-QMC-Chang2009} indicates that mean-field results provide not  only qualitative but 
quantitatively useful information in related systems.

Despite the simple nature of the mean-field approach, 
the determination of the correct ground state in the 3D lattice is far from straightforward \cite{HubHF3D-Xu-2013}.
To determine the stability of states that have 3D spatial 
dependence of the order parameter requires the use of cubic simulation cells, which quickly become 
computationally expensive as the system size increases. Additionally, 3D systems permit a wider range 
of potential ground-states, meaning the energy landscape will have more local minima and ground-state 
searches need to be increasingly thorough.
We focus on moderate interaction strengths  
($U/t\le 5$), where this approach is most reliable. 
Several strategies are employed, using large scale computations, to validate the solutions 
and the extrapolation to the thermodynamic limit.

We find that,
at high to intermediate densities, the ground state phase is of the canonical LO form
independently of interaction strength, with counter-propagating pairs and order parameter 
going to zero on a regularly spaced array of parallel planes. This is the domain in which the effect of the optical lattice is 
most apparent on the shape of the Fermi surfaces, and consequently
on the ground state phases. At low density, the Fermi surfaces become
more spherical, as they would be in the continuum, and we find that the ground state
is characterized by a superposition of pairs with non parallel momenta. In this region, where the impact of the optical lattice is less
significant and these higher-dimensional states emerge, a larger interaction
is required to induce pair 
ordering.  Systematic information is obtained on the ground-state properties, especially in the first parameter 
regime. The physical origin of the phases
and their connection to the Fermi surface topology and pairing are discussed. 

Below we first describe our computational approach in Sec.~\ref{methods}. In Sec.~\ref{sec:oplatt} the 
results for the first parameter regime, namely at high to intermediate densities, are presented, 
with discussions of the 
effects of density and polarization, and of the interaction strength. 
Results more relevant to the continuum limit, i.e., at low densities are then discussed in Sec.~\ref{2D&3Dsection}.
We conclude with a summary in Sec.~\ref{sec:summary}.

\section{Method}
\label{methods}

The starting Hamiltonian we study is,
\begin{equation}
H= - \sum_{(ij)\sigma} t_{ij} c_{i\sigma}^\dagger c_{j\sigma} -
\sum_i \left( Un_{i\uparrow} n_{i\downarrow} + \mu n_i + \frac{h}{2} m_i\right)\,,
\label{HubMod}
\end{equation}
where $c_{i\sigma}$ is a fermionic annihilation operator of spin $\sigma$ on site $i$ ,
$n_{i\sigma}=c^\dagger_{i\sigma} c_{i\sigma}$, $n_i=n_{i\uparrow}+n_{i\downarrow}$
and $m_i=n_{i\uparrow}-n_{i\downarrow}$.
In this paper we will only consider the Hubbard dispersion, i.e., $t_{ij}=t$ if $(ij)=\langle ij\rangle$ ($i$ 
and $j$ are near-neighbors) and $t_{ij}=0$ otherwise. 
The interaction will be attractive, so $U> 0$. 
Further, we will be in the regime of 
negative scattering length, since we will be concerned with $U/t \le 5$, as mentioned earlier. 
(A two-body bound state first appears at $U/t=7.91355$ for the Hubbard dispersion.)
The chemical potential $\mu$ and the ``magnetic field'' $h$ in the Hamiltonian control the
density, $n$, and the polarization, $p$. Given a supercell of $N$ lattice sites, these are defined
by $n_\sigma\equiv \sum_i \langle n_{i\sigma}\rangle/N$: $n=n_\uparrow +n_\downarrow$,
 $m=n_\uparrow -n_\downarrow$, and 
$p\equiv m/n$. 
The system is completely specified by the three parameters $U/t$,  $n$, and $p$.

Our analysis of  this Hamiltonian was performed using Hartree-Fock-Bogoliubov theory.
We transform the Hamiltonian into a diagonalizable form by employing a standard 
mean-field approximation,
\begin{align}
\sum_i &U n_{i\uparrow}n_{i\downarrow} = \sum_i Uc^\dagger_{i\uparrow} c_{i\uparrow} c^\dagger_{i\downarrow} c_{i\downarrow}\notag\\
\rightarrow \sum_i &U\Big{\{} 
\langle c^\dagger_{i\uparrow} c^\dagger_{i\downarrow}\rangle c_{i\downarrow} c_{i\uparrow} 
+ \langle c_{i\downarrow} c_{i\uparrow}\rangle c^\dagger_{i\uparrow} c^\dagger_{i\downarrow}\notag\\
&+ \langle c^\dagger_{i\uparrow} c_{i\uparrow}\rangle c^\dagger_{i\downarrow} c_{i\downarrow}
+ \langle c^\dagger_{i\downarrow} c_{i\downarrow}\rangle c^\dagger_{i\uparrow} c_{i\uparrow}
\Big{\}},
\end{align}
with constant terms omitted.

The FFLO phase is most distinctly characterized by a spatially modulated pairing 
order parameter. In order to accurately determine the relative stability of FFLO states 
with different real-space structures, we perform our calculations on simulation cells whose 
shapes accommodate those structures. The simulation cells are characterized by three basis 
vectors, $\mathbf{L}_1$, $\mathbf{L}_2$ and $\mathbf{L}_3$, whose components are integers. 
Once the cell shape is chosen we introduce Bloch states, defined as
$c_j(\mathbf{k}) \propto \sum_{\mathbf{L}} c_{j+\mathbf{L}} \exp\big[i\mathbf{k}\cdot \mathbf{L}\big]$
where $\mathbf{L}$ is a vector on the Bravais lattice having $\mathbf{L}_1$, $\mathbf{L}_2$ 
and $\mathbf{L}_3$ as basis vectors, {\em i.e.} $\mathbf{L}=l_1 \mathbf{L}_1 + 
l_2 \mathbf{L}_2 + l_3 \mathbf{L}_3$, and $\mathbf{k}$ is a vector that varies freely 
within the first Brillouin zone of the simulation cell reciprocal lattice.

Having applied the mean-field approximation, we can
use the Bloch states described above to write the Hamiltonian 
as a sum of decoupled $\mathbf{k}$-dependent pieces, $H = \sum_\mathbf{k} H(\mathbf{k})$, of the form
\begin{equation}
\begin{split}
H(\mathbf{k})= [\cm^\dagger_\uparrow \cm_\downarrow] 
\left[ \begin{array}{cc}
\Hm_\uparrow(\mathbf{k}) & \Deltam  \\
\Deltam^\dagger & -\Hm_\downarrow^T(\mathbf{G}-\mathbf{k})  \end{array} \right]
[\cm_\uparrow \cm^\dagger_\downarrow]^T
\end{split}
\label{BdGH}
\end{equation}
where $\mathbf{c}_\uparrow$ and $\mathbf{c}_\downarrow$
represent an array (row) of operators, $\{c_{i\uparrow}(\mathbf{k})\}$ and
$\{c_{i\downarrow}(\mathbf{G}-\mathbf{k})\}$
with the index $i$ running over the $N$ sites of the simulation cell.
The vector $\mathbf{G}$ is defined so that $\theta=\mathbf{G}\cdot\mathbf{L}$ is
the twist angle of the pairing order parameter after a translation by $\mathbf{L}$.
$\Hm$ and $\Deltam$ are $N\times N$
matrices with elements
\begin{equation}
\begin{split}
[\Hm_\sigma(\mathbf{k})]_{ij} & = -t_{ij}(\mathbf{k}) + \delta_{ij} ( D_{i\sigma} - \mu -s_\sigma h/2) \\
[\Deltam]_{ij}    & = \delta_{ij} \Delta_i .
\end{split}
\end{equation}
In the above equation $t_{ij}(\mathbf{k})=\sum_\mathbf{L} \exp(i\mathbf{k}\cdot\mathbf{L}) 
t_{i,j+\mathbf{L}}$, $s_{\uparrow/\downarrow} = \pm 1$
and $D_{i\sigma}$, $\Delta_i$, $\mu$ and $h$ are determined by the
requirement that the Free energy 
$F = \langle H\rangle  - T S$
is a minimum for the target average densities ${n}_\sigma$.  All of our calculations
are performed at $T/t=0.01$. 
This amounts to the 
following self-consistency equations
\begin{equation}
\begin{split}
D_{i\sigma} &=  -U \int d\mathbf{k} \langle c^\dagger_{i\sigma'}(\mathbf{k})c_{i\sigma'}(\mathbf{k}) \rangle \\
\Delta_i        &= -U \int d\mathbf{k} \langle c_{i\,\downarrow}(\mathbf{k})c_{i\,\uparrow}(\mathbf{k}) \rangle\\
n_\sigma  &=  N^{-1} \sum_i \int d\mathbf{k} \langle c^\dagger_{i\sigma}(\mathbf{k})c_{i\sigma}(\mathbf{k})\rangle.
\end{split}
\label{gapEq}
\end{equation}
where in the first equation $\sigma'$ is the opposite of $\sigma$.

We make the following {\em initial} ansatz for the spatial form of the order parameter,
\begin{equation}
\Delta_i^{(0)} = \sum_\qm \Delta^{(0)}_\qm e^{i\qm\cdot \rb_i}.
\label{pairing_ansatz}
\end{equation}
This represents a summation of plane wave modes
characterized by a set of symmetry-related pairing vectors $\qm$. The spiral (FF) phase 
corresponds to a single $\Delta^{(0)}_\qm\ne0$ or, in real space, to $\Delta^{(0)}_i\propto e^{i\qm\cdot \rb_i}$. The 
linear (LO) phase has $\Delta^{(0)}_{\pm\qm}\ne0$
with $\qm\propto (0,0,1)$, $(0,1,1)$, or $(1,1,1)$
and $\Delta^{(0)}_i\propto \cos(\qm\cdot \rb_i)$.
In addition, we consider
2D structures of the form
$\Delta_i \propto \cos(\qm_y\cdot \rb_i)+\cos(\qm_z\cdot \rb_i)$,
with $\qm_y=|\qm|(0,1,0)$ and $\qm_z=|\qm|(0,0,1)$,  
and 3D structures of the form 
$\Delta_i \propto \cos(\qm_x\cdot \rb_i)+\cos(\qm_y\cdot \rb_i)+\cos(\qm_z\cdot \rb_i)$,
with $\qm_x=|\qm|(1,0,0)$ and $\qm_y$, $\qm_z$ as before. 

\begin{figure}
\includegraphics[width=\columnwidth]{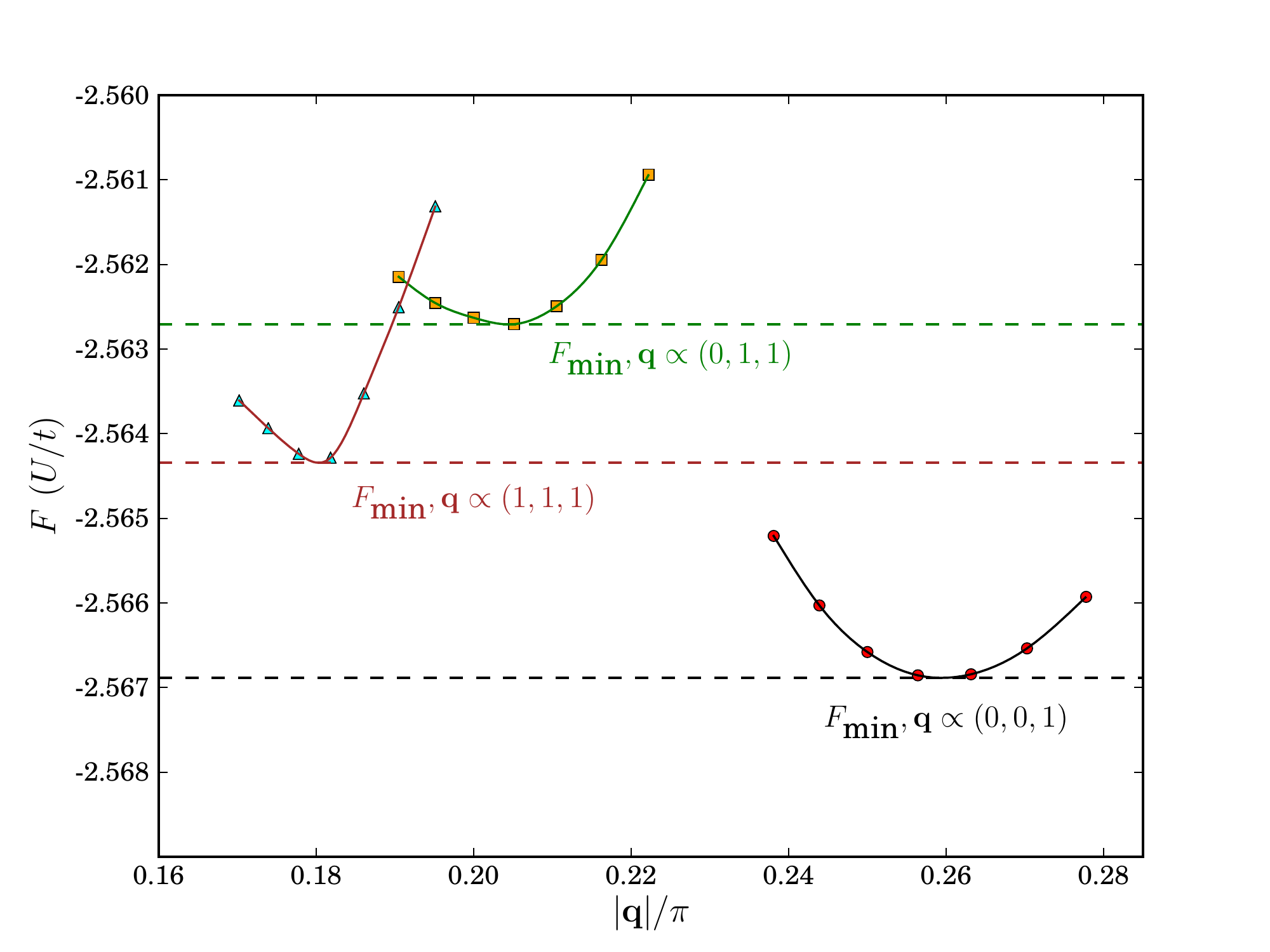}
\caption{Determining the nature of the FFLO state.
The  free energies of linear pairing-wave states with $\qm\propto (0,0,1), (0,1,1)$, and 
$(1,1,1)$ are compared for $n=0.76$, $p=0.23684$ at $U/t=5.0$.
Here a scan over $|\qm|$ has been performed to determine 
the optimal $|\qm|$ and the corresponding minimum free energy, which is indicated for each $\qm$-direction by the 
dashed line. In this case the ground state has $\qm\propto (0,0,1)$.
}
\label{freeEvsq}
\end{figure}

Our procedure allows us an unbiased search of the ground state within the general form of the 
candidate orders which are tested.
Different choices of $\Delta_i^{(0)}$ determine different shapes of the simulation cell which, in turn, constrain
the form of the self-consistent $\Delta_i$. 
A typical example, for a linear phase, might have $\mathbf{L}_1=(1,0,0)$,  $\mathbf{L}_2=(0,1,0)$, 
and $\mathbf{L}_3=(0,0,50)$. 
After the shape of the simulation cell has been selected, we 
perform a scan over $|\qm|$ to determine the optimal $|\qm|$ corresponding to the minimum energy 
ground state for each $\qm$-direction (or for the higher-dimensional structures, the minimum energy for 
each set of $\qm$'s). 
For each calculation in the scan, we sum over a sufficiently dense ${\mathbf k}$-grid to remove all
finite-size effect except for the constraint on the form of the order from the shape of the simulation cell.
In the case above, for example, our calculation would use a ${\mathbf k}$-point grid 
which has dimensions of a few in the $\mathbf{L}_3$ direction and a few
hundred in the $\mathbf{L}_1$ and $\mathbf{L}_2$ 
directions.
This technique allows the calculation to accommodate the spatial modulation of the phase and 
approach the thermodynamic limit.

This procedure is sketched schematically for linear phases in Fig.~\ref{freeEvsq}. 
The calculations are to determine the true ground state among pair-ordered states with pairing vector $\qm$ directed along 
either $(1,0,0), (1,1,0),$ or $(1,1,1)$. 
For each $\qm$-direction we perform a scan to determine the optimal
$|\qm|$, varying the simulation cell size to ensure that it is commensurate with the targeted value of $|\qm|$.
To rule out orders other than linear, we carry out searches for the 2D and 3D structures described above.
Further, we 
increase the simulation cell size in directions other than ${\mathbf q}$ to verify the stability of the solution.

\section{Optical lattice regime}
\label{sec:oplatt}

\begin{figure}
\includegraphics[width=0.85\columnwidth]{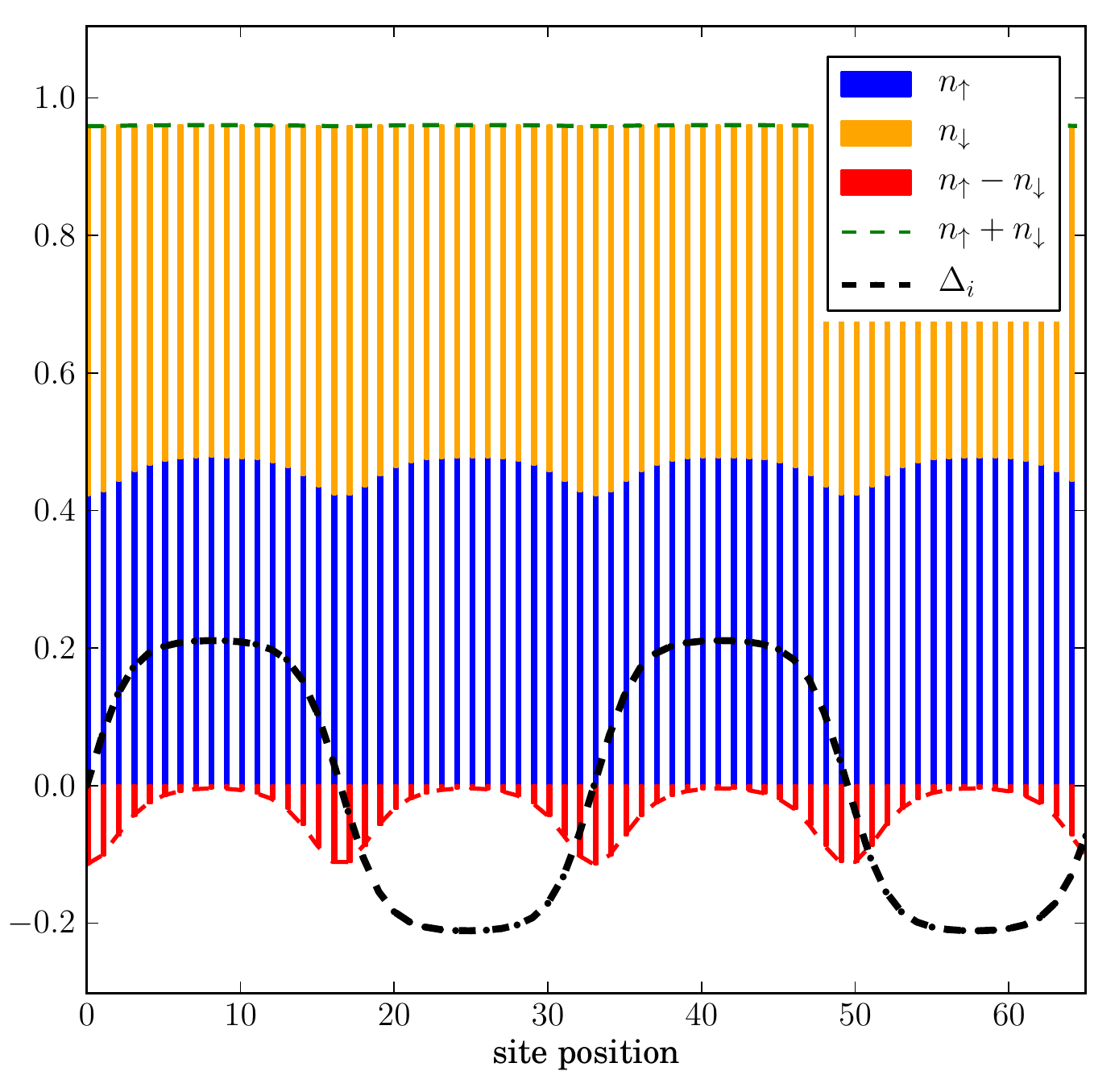}
\caption{Illustration of the real-space properties of the LO state.
Shown is the ground state at $U/t=3.0$, with $n=0.96$ and
$p=-0.041667$. (The $\downarrow$-spin is chosen to be the majority spin in this work.) 
The densities are plotted as a stacked bar chart, with the
total density indicated by the green dashed line. The difference between
the densities, the spin density, is plotted in red. The pairing order
parameter is plotted in black.
The domain wall character 
of the pairing wave is evident here and the amplitude of the order parameter is large.} 
\label{U3.0_dens0.96_sp0.02_realspace}
\end{figure} 

We first consider the region of high to intermediate densities and low polarizations,
where the characteristics of the ground-state phases of the system are significantly impacted
by the presence of the optical lattice. This effect is most clearly reflected in the shape
of the Fermi surface. At high density the Fermi surfaces of both spin species are very distinct 
from their spherical counterparts in the continuum. 
The nature of the pairing mechanism and its 
connection to the shape of the Fermi surfaces is further discussed below.

As described in Sec.~\ref{methods}, the set of pairing wave vectors that leads to the 
minimum energy state determines the spatial structure of the pairing order parameter of 
that state. We find that in the optical lattice regime the system favors states with two $\qm$ 
vectors, which results in an order parameter that is a linear pairing wave. 
The spiral state is energetically less favorable and never found to be the ground state 
in the regime we have investigated. This is similar to the situation in 2D \cite{2Dattractive}
and is consistent with the results from the 3D repulsive Hubbard model  \cite{HubHF3D-Xu-2013}
after particle-hole mapping. 
The properties of 
the linear phases, including the direction of the $\qm$ vectors, exhibit dependence on density 
and polarization, and will be discussed in detail in Sec.~\ref{ssec:oplat-np}.

In Fig.~\ref{U3.0_dens0.96_sp0.02_realspace}, we present a characteristic example of the linear LO 
phase,
in order to illustrate some of its real space properties. 
The ground state at these parameters is found to have $\qm\in\{|\qm|(0,0,1), |\qm|(0,0,-1)\}$. 
At small polarizations and high densities such as this particular case, the domain wall nature of the pairing wave is evident. 
The densities of both spin species exhibit spatial modulation, with the density of the majority equal to 
the density of the minority at the peak of the order parameter. The greatest difference between the 
minority and majority density occurs at the nodes of the order parameter. This results in a peak of the 
spin density, which can be understood as the localization of excess spin at the nodes of the order 
parameter. The quantity $\alpha \equiv m\,\pi/|{\mathbf q}|$ characterizes the total density of the excess spin 
within each nodal region (a stack of planes perpendicular to ${\mathbf q}$). 
The overall charge density of the system is essentially a constant in this case.

\begin{figure}
\includegraphics[width=0.8\columnwidth]{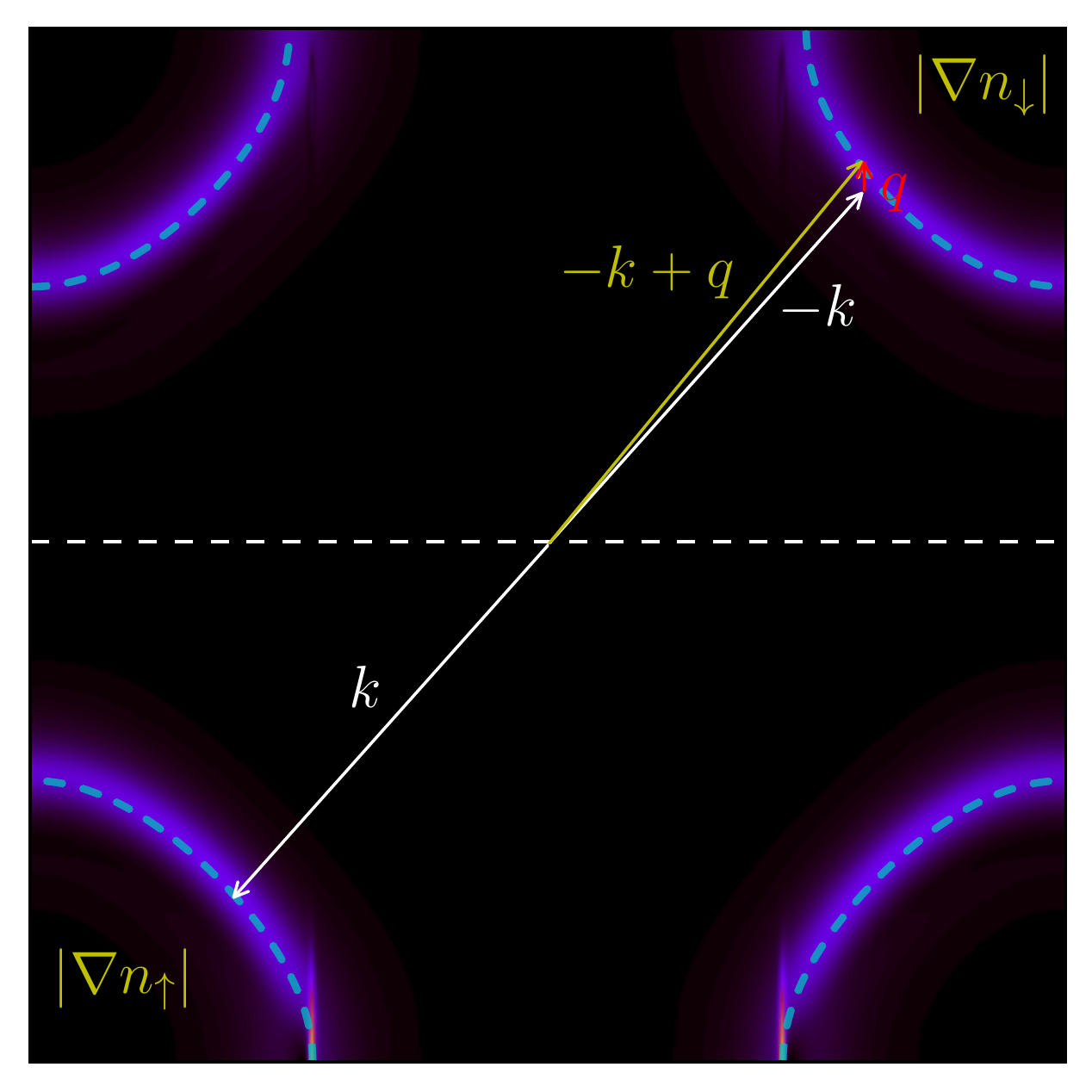}
\caption{Momentum-space properties of the ground-state at $U/t=3.0$ with $n=0.96$ and
$p=-0.041667$. Above the white dashed line is the top half of the Fermi surface of
the majority spin species $(\downarrow)$, and below is the bottom half of the 
minority $(\uparrow)$  Fermi surface, for a 2D slice in the $k_x$-$k_z$
plane at $k_y=0$. 
The non-interacting Fermi surfaces are plotted using a dashed blue line.
The interacting Fermi surfaces have a similar shape, slightly modified from the 
non-interacting ones, so a collection of pairs can form with
a common $\qm$ (drawn in red) by the $\mathbf{k}\rightarrow-\mathbf{k}+\qm$ construction. The sharp segments of the 
Fermi surface indicate regions where electrons have not paired.
}
\label{U3.0_dens0.96_sp0.02_kspace}
\end{figure}

The momentum-space properties of the same state are plotted in 
Fig.~\ref{U3.0_dens0.96_sp0.02_kspace} using the gradient of the momentum distribution. 
This quantity gives the position of the underlying Fermi surface which, as shown later (Fig.~\ref{U5.0_kspace_props_dens0.60_sp0.09_sp0.17}),
need not coincide with the non-interacting one.
Illustrated on the plot is the 
pairing construction, $\mathbf{k}\rightarrow-\mathbf{k}+\qm$, by which electrons near the 
Fermi surfaces of the two different spin species form pairs with finite 
momentum $\qm$. 
In this case, a slight modification of the shape of the interacting Fermi surfaces from 
the non-interacting ones allows electrons along large sections of both Fermi surfaces
to form pairs with a single pair of $\qm$'s with common magnitude $|\qm|$. 
The resulting order parameter is a sum of plane waves, whose collective interference 
serves to lower the energy of the state and produce the standing wave structure visualized
in Fig.~\ref{U3.0_dens0.96_sp0.02_realspace}. For the set of parameters corresponding to 
the state in the figure, and the slice of momentum-space plotted, a large fraction of the
Fermi surface is smeared as a consequence of pair formation. The sharp features at the 
bottom of the minority Fermi surface identify a region where the Fermi surface is still intact and
remains un-gapped. This is consistent with $\alpha\ne 1$ and a metallic nodal region
\cite{2Dattractive}. In this case, the intact portion of the minority Fermi surface is small, indicating 
that most of the electrons near the Fermi surface have paired.

Having highlighted the important features of the FFLO phase in
the optical lattice regime, in both real and momentum space, we will
now discuss in more detail the effect of density, polarization, and interaction strength 
on these features. A final phase diagram summarizing all our calculations 
is then presented in Sec.~\ref{ssec:opat-U}.

\subsection{Density and polarization}
\label{ssec:oplat-np}

\begin{figure}
\includegraphics[width=\columnwidth]{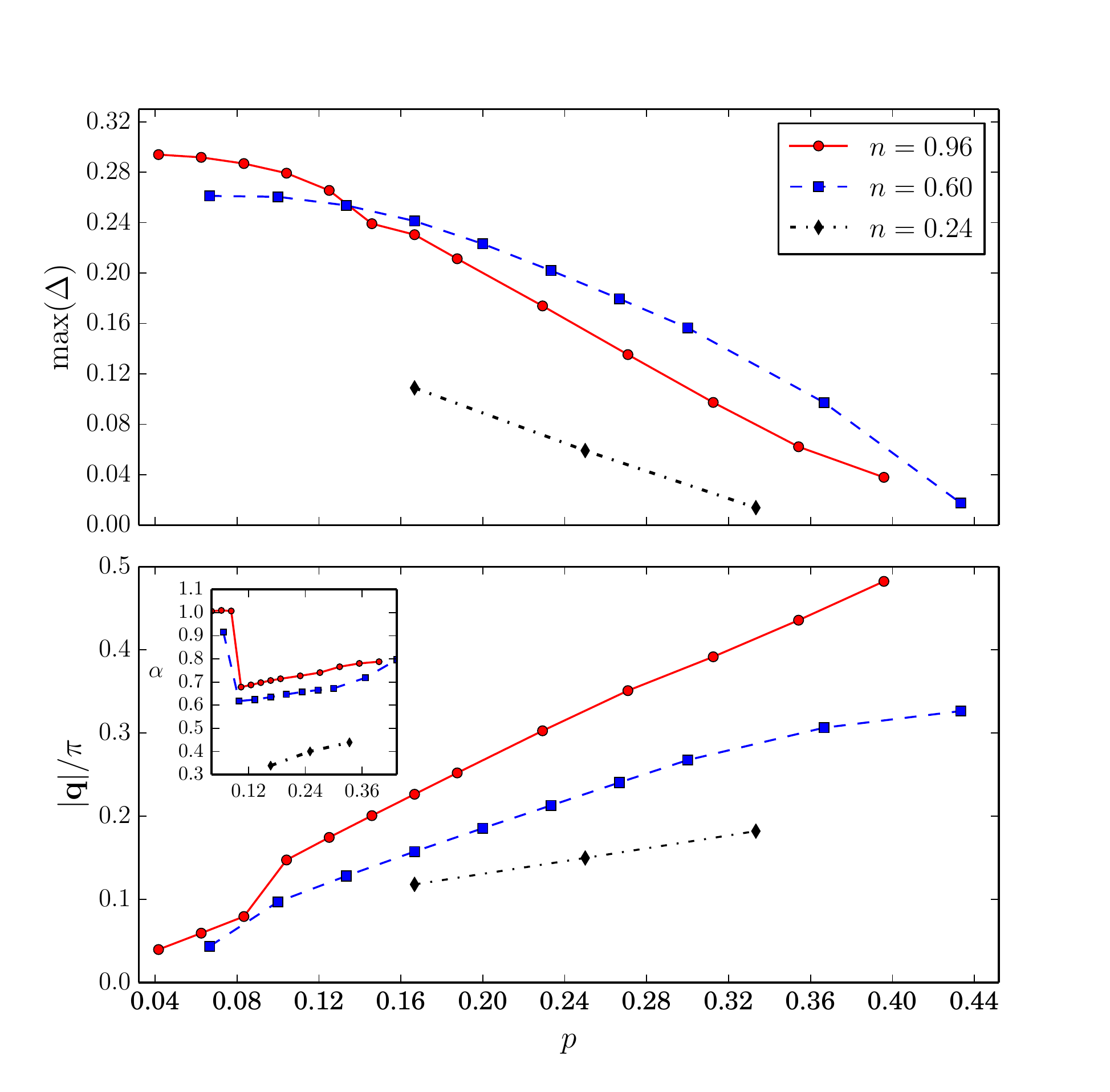}
\caption{Plot of $\textrm{max}(\Delta)$ and $|\qm|/\pi$ versus polarization for $n=0.96$, 
$0.60$, and $0.24$ at an interaction strength of $U/t=4$. The transition from a ground state with 
$\qm\propto (1,1,1)$ to one with $\qm\propto(0,0,1)$ can be seen around $p=0.08$ for $n=0.96$ and $n=0.60$
where the value of $\alpha$ (inset) drops dramatically.
}
\label{U4.0_dens0.96_dens0.60_dens0.24_orderp_q_slice}
\end{figure}

In this section we  
examine in further detail the characteristics of the ground-state phases as they depend on density 
and polarization. 
At each selected interaction strength $U/t$, we map out the complete $n$-$p$ phase diagram.
The behavior of the linear phase as a function of
polarization,  for $n=0.96$, $n=0.60$ and $n=0.24$ at $U/t=4$ is illustrated in 
Fig.~\ref{U4.0_dens0.96_dens0.60_dens0.24_orderp_q_slice}. At large polarizations, 
near the onset of pairing order, the order parameter is small, large portions of the Fermi surfaces of the two spin species 
remain ungapped, and those that are gapped remain sufficiently sharp to be precisely located.
As the polarization decreases, pairing is enhanced and the pairing order parameter increases as expected. 
Lower polarization is also where it is more likely to have $(1,1,1)$ order, and a transition to it 
from $(1,0,0)$ can be seen in the figure where the value of $\alpha$ decreases significantly,
for $n=0.96$ and $n=0.60$. The appearance of  $(1,1,1)$ order involves larger Fermi surface reconstructions, 
in a way similar to the nesting mechanism for the formation of spin-density waves in the 3D repulsive case 
\cite{HubHF3D-Xu-2013}.

\begin{figure}
\includegraphics[width=\columnwidth]{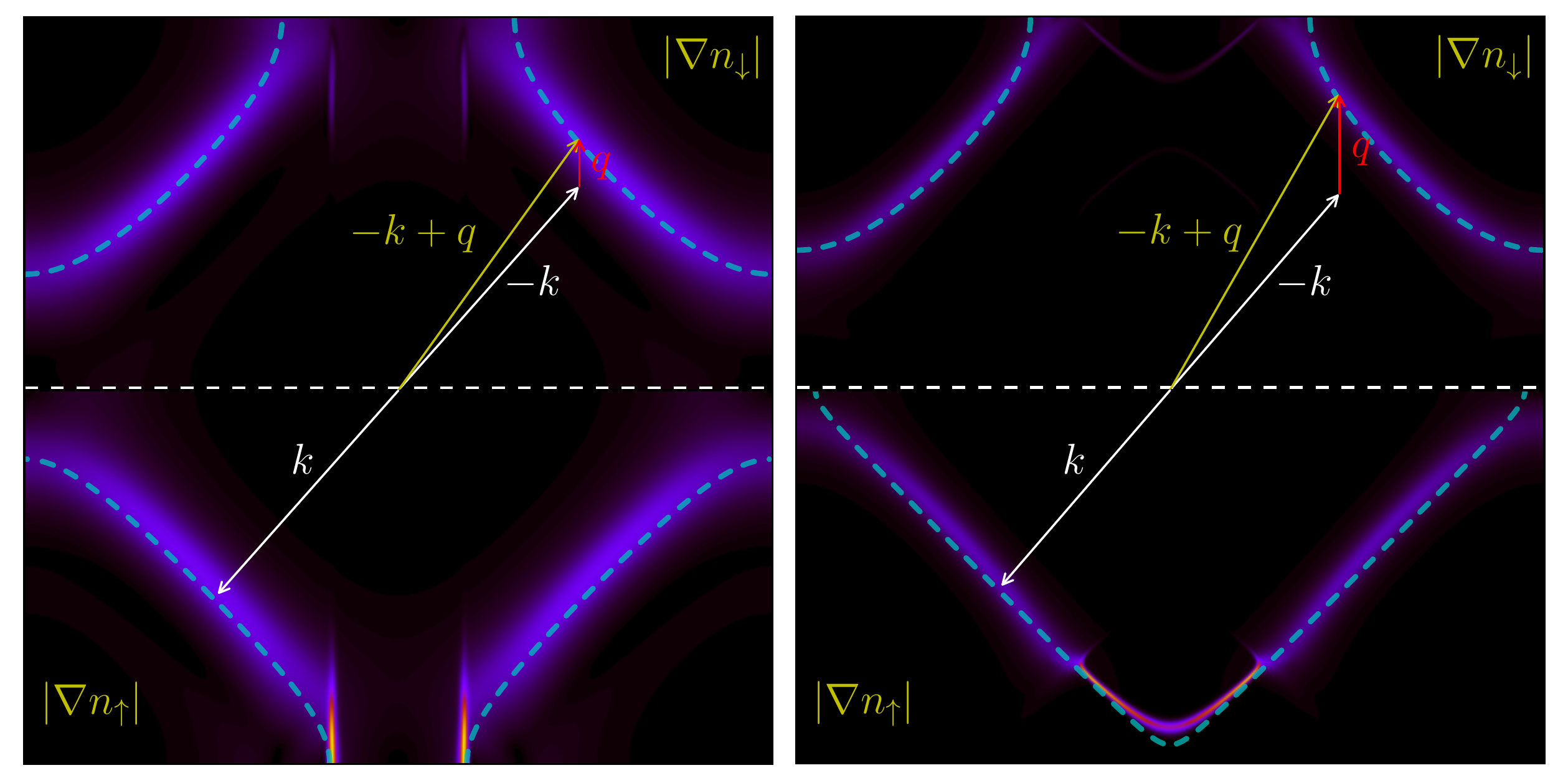}
\caption{Comparison of the momentum-space properties of the linear phase at $U/t=4$, $n=0.60$, 
for $p=-0.13333$ (left) and $p=-0.3$ (right).
At large polarization the system requires
a large $|\qm|$ to form electron pairs. 
The modification of the interacting Fermi surfaces from the non-interacting ones
is very apparent in the right panel. This modification allows more electrons
to participate in pairing. As the polarization decreases the non-interacting Fermi 
surfaces of the two species become closer in size and more similar shape, so 
pairing can occur with less modification of the interacting Fermi surfaces and a 
smaller $|\qm|$.}
\label{U5.0_kspace_props_dens0.60_sp0.09_sp0.17}
\end{figure}

\begin{figure}
\includegraphics[width=\columnwidth]{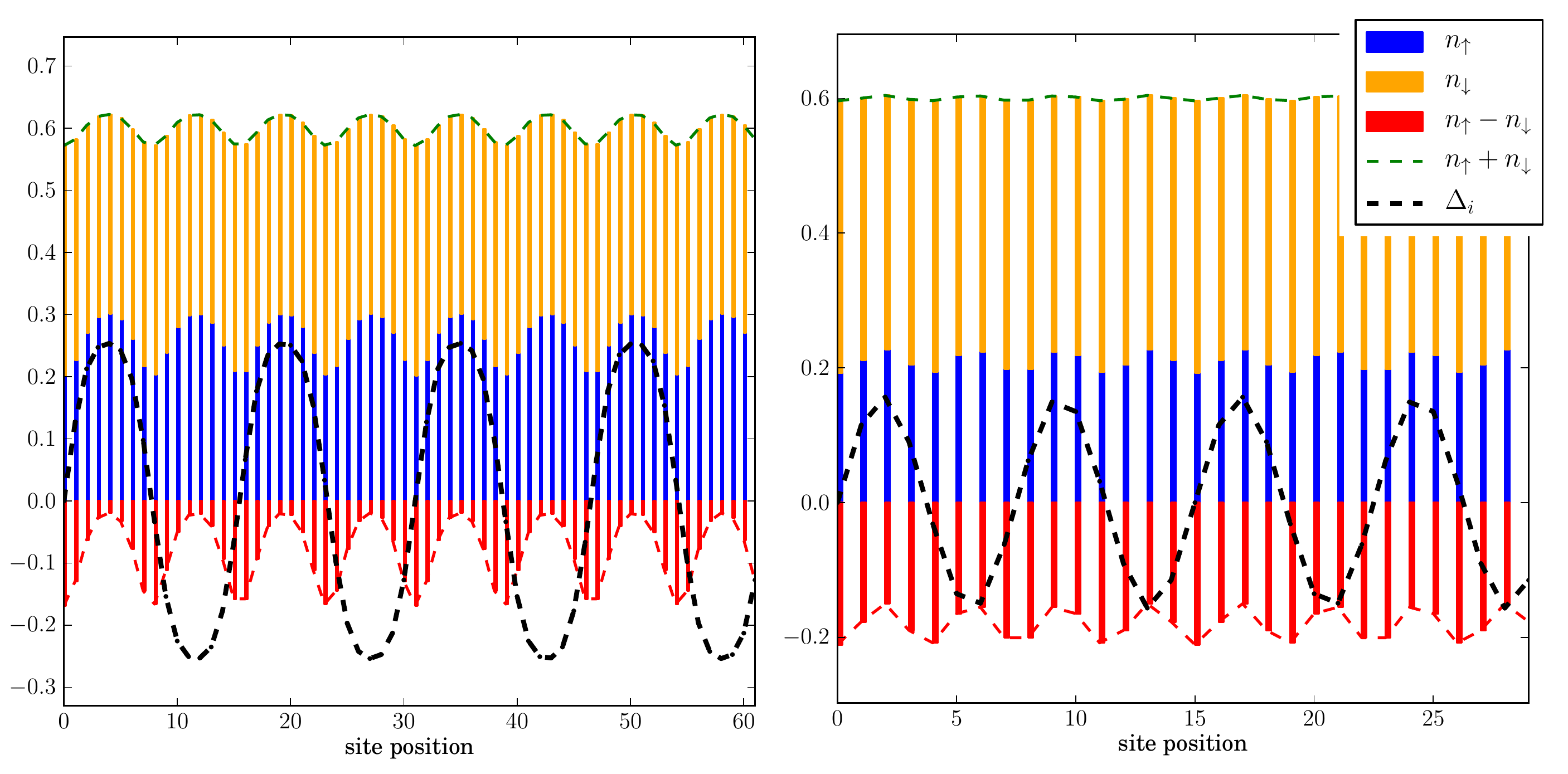}
\caption{Comparison of the real-space properties of the linear phase at $U/t=4$, $n=0.60$, 
for $p=-0.13333$ on the left and $p=-0.3$
on the right. At small polarization the pairing wave has domain walls, with
localized excess spin. As the polarization increases the pairing wave becomes sinusoidal
and the amplitude decreases.
}
\label{U4.0_dens0.60_sp0.02_sp0.09_real_space}
\end{figure}

Figures~\ref{U5.0_kspace_props_dens0.60_sp0.09_sp0.17} 
and \ref{U4.0_dens0.60_sp0.02_sp0.09_real_space} 
visualize and compare the momentum- and real-space properties, respectively, for different values of the polarization.
As already discussed, the underlying Fermi surface of the LO phase 
can deviate from the non-interacting one. 
The numerical solution can be understood by the momentum space nesting caused by the surface reconstruction
and the pairing mechanism that ensues. 
At large polarizations, 
a larger ${\mathbf q}$ is required, and smaller portions of the Fermi surface can support pairing, hence 
weaker order parameter. 
Eventually, as one moves farther from the transition and deeper into the LO phase, the Fermi surface is heavily smeared,
the order parameter comprises many (collinear) momenta. Correspondingly, in real space 
the order parameter remains purely sinusoidal, the density modulation is weak, and the 
excess spin is \emph{not} localized at large polarization. As the polarization decreases, 
the physics is better understood in the language of weakly interacting
domain walls, with the excess spin more localized at the nodes of the order parameter, and strong density 
modulation.

Figure~\ref{U4.0_dens0.96_dens0.60_dens0.24_orderp_q_slice} also captures
the behavior of the ground state properties as a function of density. At high densities the 
presence of the underlying lattice has a significant effect on the shape of the Fermi surface. 
For states at high density $|\qm|$ is large compared to states at the same polarization but lower 
density. Additionally, the effect of polarization on $|\qm|$ is more prominent at higher density, where
a larger spin imbalance is required to achieve the same polarization
than is required at a lower density. This effect can be seen, for example, by 
comparing the slopes of $|\qm|/\pi$ vs.~$p$ for $n=0.96$ and $n=0.24$. The slope of the $n=0.96$ curve 
is significantly steeper than the slope of the $n=0.24$ curve. Also, all values of $|\qm|$ are smaller for 
$n=0.24$ compared to $n=0.60$ and $n=0.96$, which reflects the smaller mismatch between 
Fermi surfaces at lower density. 

\subsection{Interaction strength}
\label{ssec:opat-U}

 \begin{figure*}
\includegraphics[width=\textwidth]{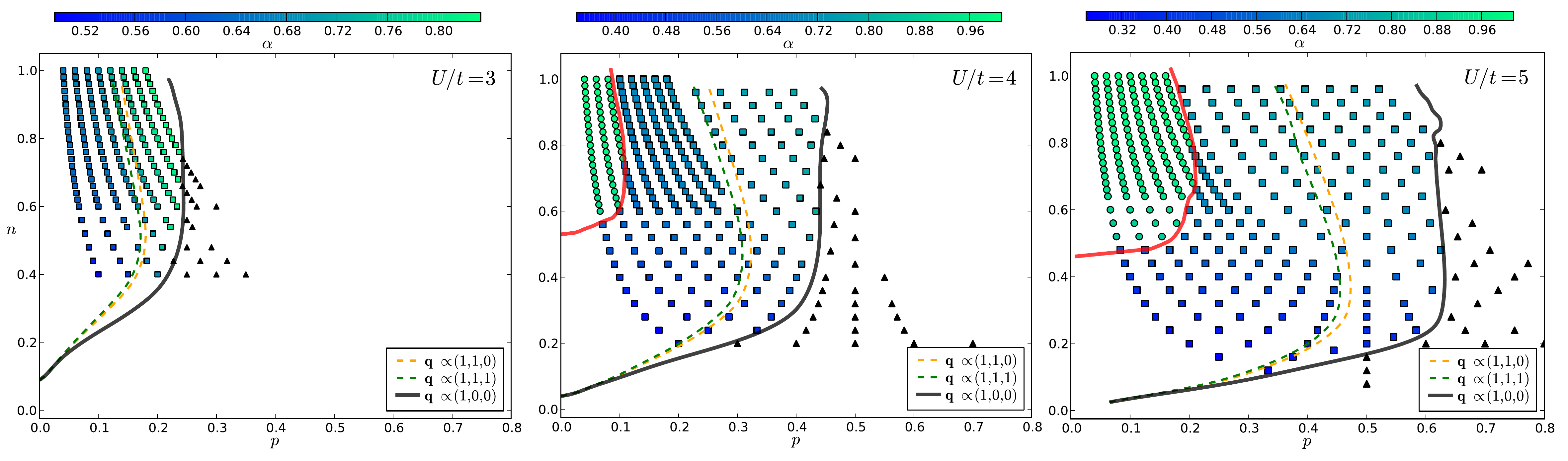}
\caption{Density-polarization phase diagrams at several values of interaction strength.
Circles indicate a solution with $\qm\propto (1,1,1)$, and squares indicate $\qm\propto(0,0,1)$. The black 
triangles represent a solution without order. The color scale gives the value of $\alpha = m\pi/|\qm|$ (note
that this scale is different for the three diagrams). The solid black and red lines represent phase boundaries.
The black lines indicate the transition from an unordered state to an ordered state with $\qm\propto(0,0,1)$, and the 
red lines in the right two panels indicate the transition from an ordered state with $\qm\propto(0,0,1)$ to one with 
$\qm\propto(1,1,1)$. The dashed orange and green lines are the estimates (see text) of the regions where the 
system could support an ordered solution with $\qm$ in the given direction.}
\label{phase_diagrams_horizontal}
\end{figure*}

In Fig.~\ref{phase_diagrams_horizontal} we summarize the phase diagrams for three values of 
interaction strength.
The interaction strength plays a significant role in determining the stability 
of pair ordered ground-state phases. The LO ground state becomes more stable as the 
polarization decreases and the interaction strength increases. This behavior is evident in the 
phase diagrams for $U/t=3$, $U/t=4$ and $U/t=5$,
which show that the area of phase space occupied by an ordered state grows larger with 
increasing interaction strength. The trend suggests that as the Fermi surfaces of the two spin 
species become closer and more similar in shape, pairing order becomes increasingly 
energetically favorable. This is especially true at higher interaction strengths, which allow a 
more significant reshaping of the Fermi surface to improve nesting and permit a
larger number of electrons to participate in pairing. 

\begin{figure*}
\includegraphics[width=\textwidth]{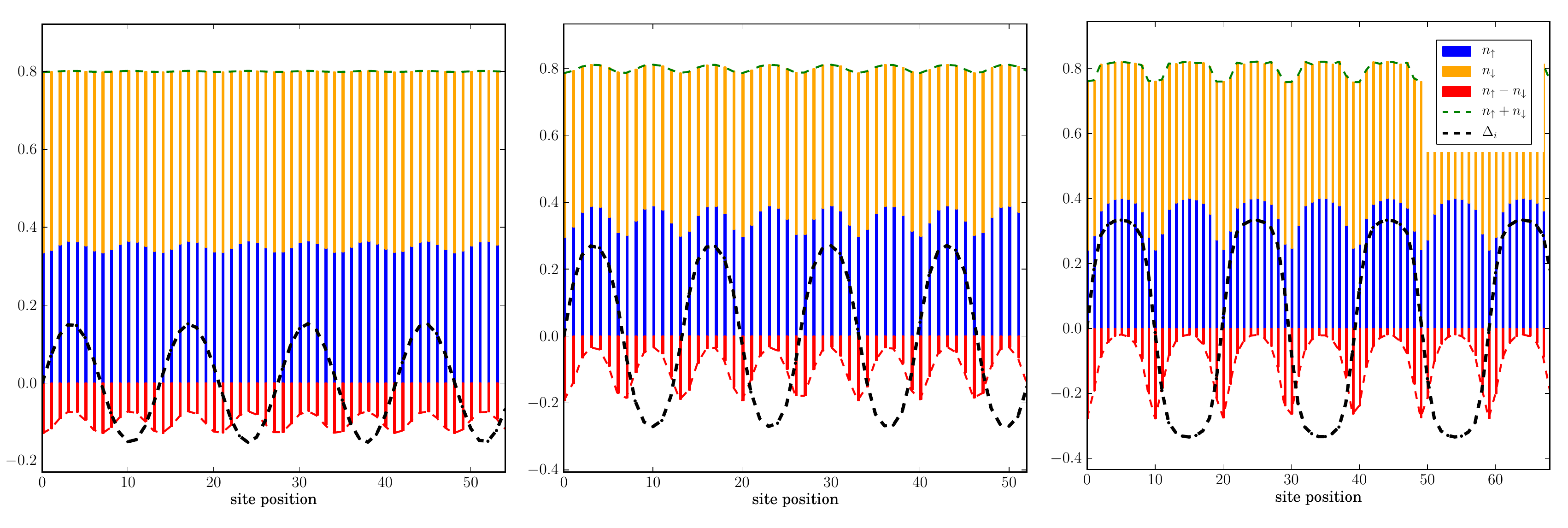}
\caption{Comparison of real-space properties at $p=-0.125$, $n=0.80$ for $U/t=3,4,$ and $5$ 
(from left to right). As the interaction strength increases the pairing wave begins to develop domain
walls, and the amplitude increases. Additionally, the excess spin becomes more strongly localized 
at the nodes of the order parameter, and the density modulations grow. At $U/t=3,4$ the state has
$\qm\propto (0,0,1)$. At $U/t=5$ the state has $\qm\propto (1,1,1)$, and $|\qm|$ increases (note
the different cell size in the right panel from the other two).}
\label{U3_U4_U5_realspace_m0.05_dens0.80}
\end{figure*}

In each of the phase diagrams in Fig.~\ref{phase_diagrams_horizontal}, we show estimates of the 
parameter regions in which a local minimum exists for a pair-ordered state with pairing vector $\qm$ 
directed along either $(1,0,0), (1,1,0),$ or $(1,1,1)$. These estimates are obtained by calculating $U/t$ 
from the gap equation for $\Delta=0$ at fixed $n$ and $p$ with $\qm\propto (1,0,0), (1,1,0),$ or $(1,1,1)$. 
For each $\qm$-direction we perform a scan in $|\qm|$ to determine the minimum $U/t$ required to induce 
pairing at the chosen $n$ and $p$. We repeat this procedure for several hundred sets of $n$ and $p$, which 
provides a map of the critical $U$ across the phase-space. For a given $U$ and $\qm$-direction this defines 
a curve in $n$ and $p$ outside of which the system will not have a pair-ordered solution with 
a pairing vector in the given $\qm$-direction. These curves are indicated 
for the different $\qm$-directions on the phase diagrams. They help guide our survey of the
density-polarization phase space by indicating which states (defined by the direction of $\qm$) 
to consider in the fully self-consistent calculations. We then perform the numerical procedure outlined 
in Sec.~\ref{methods}, and sketched in Fig.~\ref{freeEvsq}, which determines the true ground-state from the 
stable pair-ordered states. It is the full numerical search, the results of which are represented by the symbols in 
the phase diagrams, that provides the actual form of the order at each point.

In addition to affecting the overall stability of pair ordered states relative to uniform states,
the interaction strength also affects the density and polarization dependence of the transitions 
between the ordered phases, which are characterized by different sets of $\qm$ vectors. At $U/t=3$, 
we find that linear pairing order with the pairing-wave vector $\qm$ directed along the $(0,0,1)$-direction 
is the ground-state for all values of density and polarization. We found no region of the $U/t=3$ phase 
diagram in which the commensurate phase, defined by a density of one excess particle per node of the 
order parameter, is stable. This is seen in the $U/t=3$ phase diagram, where no symbol 
reaches the color for $\alpha=1$. Instead, at low polarization and near half-filling $\alpha$ approaches 2/3.
This behavior is caused by the nature of the LO ground-state at $U/t=3$, which has $\qm$ directed
along the $(0,0,1)$-direction with $|\qm|\neq m\pi$. We observe that the commensurate phase 
has $\qm\propto (1,1,1)$, which does occur for $U/t=4$ and $U/t=5$.   

At $U/t=4$, a transition occurs between the linear phases with $\qm \propto (0,0,1)$
and $\qm \propto (1,1,1)$. The diagonal phase ($\qm \propto (1,1,1)$) occupies
the high to intermediate density and low polarization region of the phase space. In a portion of
this region the commensurate phase is stable. At intermediate to high polarization, or for sufficiently
low density, the pairing wave is directed along $\qm \propto (0,0,1)$ and the state is no longer 
commensurate.

The behavior at $U/t=5$ is similar to that at $U/t=4$, but with a 
larger region of stability  for the diagonal phase. Again, in a portion of this region the commensurate phase is 
stable. As with $U/t=4$, at large polarizations or low densities, the pairing wave is directed along $(0,0,1)$,
occupying a large portion of the phase space. The $(0,1,1)$-order is predicted by the gap equation
to be stable in a large region but is never the true ground state.

The effect of increasing interaction strength is also apparent in the real-space character of the phases.
This effect is visualized in Figure~\ref{U3_U4_U5_realspace_m0.05_dens0.80}.
As interaction strength increases the pairing wave develops domain walls and the 
amplitude of the pairing wave and the density modulations grow. The larger density modulations cause 
the peaks of the spin density to become sharper, making the excess spin more localized.  
 
\section{Approach to the continuum: trapped Fermi gases}
\label{2D&3Dsection}

\begin{figure}
\includegraphics[width=\columnwidth]{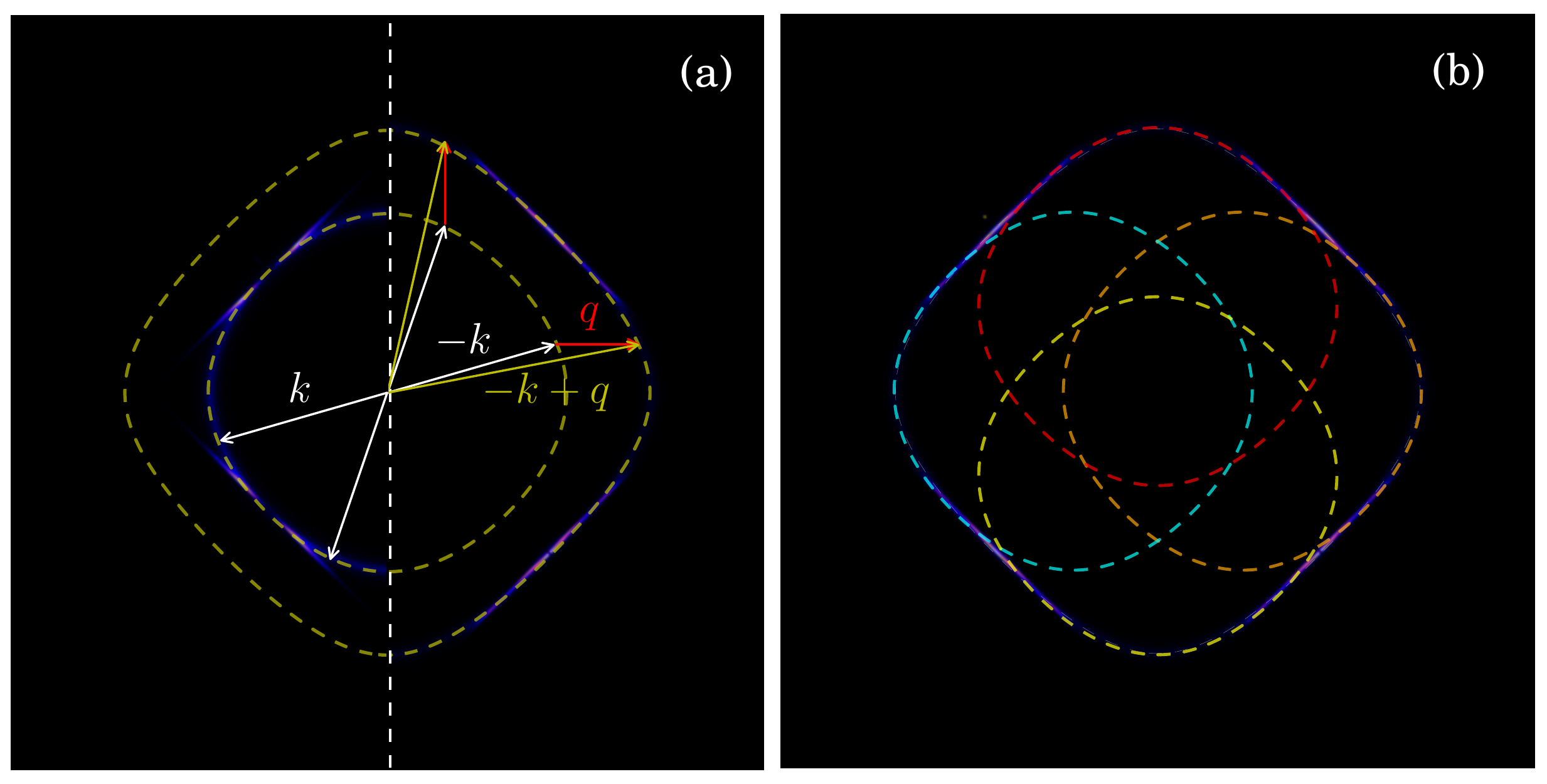}
\caption{Illustration of pairing mechanism for an ordered phase with 2D modulations. 
The system has parameters, $n=0.18$, $p=-0.4444$ and $U/t=5$.
(a) The spectral function is shown for the minority (left half) and majority (right half) spins. 
The corresponding non-interacting Fermi surfaces are indicated by the dashed yellow lines
(the majority on the left and the minority on the right). 
Two pairing wave vectors are illustrated. The reflection of each about the origin will lead to 
$-{\mathbf q}$.
(b) The majority spectral function is overlaid with the non-interacting majority 
Fermi surface (white solid line). 
Each dashed curve represents the non-interacting minority Fermi surface 
translated by one of the ${\mathbf q}$ vectors. The bright sections of the spectral function, indicating 
unpaired regions of the Fermi surface, coincide with the sections that are not overlain by the 
translated minority Fermi surface.
}
\label{fs_2D_U5.0_n0.18_m0.04_v3}
\end{figure}

At low density the effect of the lattice on the shape of the Fermi surface is less significant
and the properties of the system begin to resemble those of fermions in the continuum. 
In order to describe the experimental situation of trapped atomic gases, the Hamiltonian we have 
been using can be thought of as a discretized representation of the continuum \cite{Carlson-FG-2011-PRA}.  
The calculations must then
be at the extremely dilute limit, with large supercells, 
to obtain realistic results in the thermodynamic limit in this situation. 
 This is not the focus of the present study. However, we do extend our optical lattice studies above to 
selected lower densities. The results shed light on the approach to the continuum limit, which we discuss 
briefly here.

In this region, at large
polarizations, we find that phases with a larger set of $\qm$'s become energetically 
favorable relative to linear phases, which have just a single pair of $\qm$'s. 
An example is illustrated in Fig.~\ref{fs_2D_U5.0_n0.18_m0.04_v3}, which
plots slices of the Fermi surfaces and spectral functions, 
and sketches the pairing construction, for a 2D state. The system forms pairs with 
$\qm=\pm|\qm|(0,0,1)$ and $\qm=\pm|\qm|(0,1,0)$, as compared to 
the case of linear order where pairs can form only with $\qm=\pm|\qm|(0,0,1)$. The additional $\qm$'s 
allow for more pairing, again at little cost in kinetic energy, which lowers the total energy
of the state. 

As depicted in the right panel, favorable nesting is
achieved with four pairing wave vectors, which
allows nearly every section of the majority spin surface to be 
 covered by the minority surface. 
The sections that are not covered remain as bright spots, 
because the electrons 
in those regions have not paired and the Fermi surface remains intact. 

\begin{figure*}
\includegraphics[width=\textwidth]{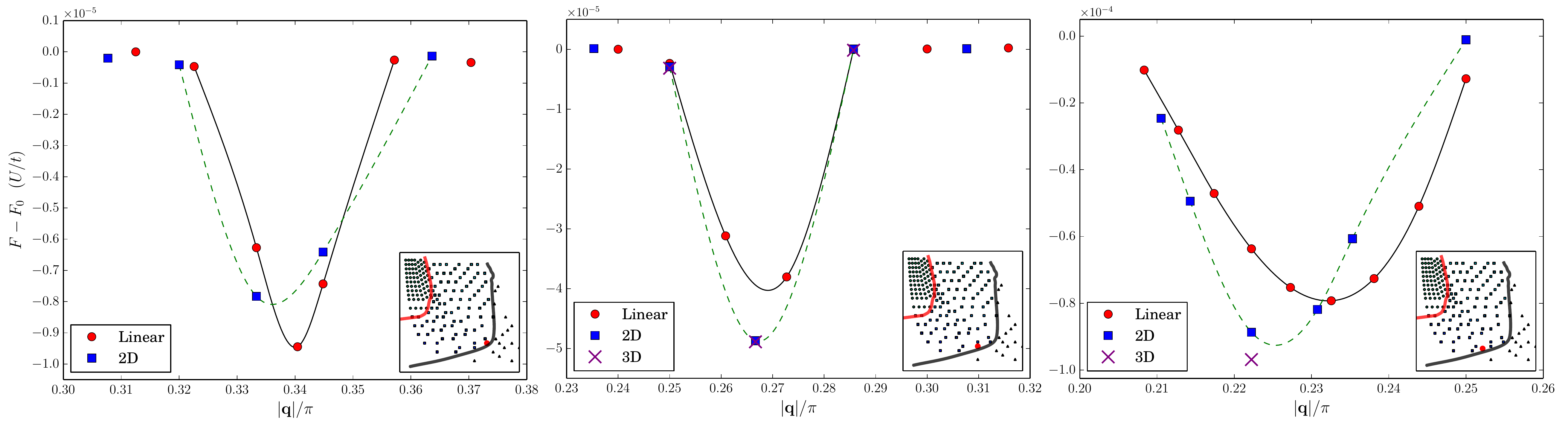}
\caption{Plots of free energy vs.$\,|\qm|$ corresponding to the points indicated by a red circle on the phase diagram (inset). 
In each case the free energy is shifted by the free energy of the uniform state at those parameters, $F_0$. The curves represent 
a fit, performed using a cubic spline interpolation scheme, to determine the minimum free energy for each state. Proceeding from left to right, these plots illustrate the emergence of higher-dimensional ground states as the lowest energy 
ground states of the system as the density and polarization are decreased near the onset of pairing order. }
\label{3D_emergence_wlabels}
\end{figure*}

In the dilute Fermi gas limit, the Fermi surfaces will be spherical and will not retain the features in the 
example above which made a 3D structure more favorable. However, more wave vectors can be involved 
which can create a more complicated structure of modulation to lower the interaction energy. This 
situation is seen in the electron gas, in which complex structures of spin-density waves are the 
true ground state in Hartree-Fock theory \cite{Zhang-PRL-HF-HEG,Bernu-PRL-HEG-HF}. Here we show 
one example, in Fig.~\ref{3D_emergence_wlabels}, of 
the emergence of phases with higher dimensional spatial variation of 
the order parameter as the lowest energy ground states of the system. 
At $n=0.24$, $p=0.5833$ the linear solution
has the lowest energy. However, moving to lower density and polarization, but still near 
the onset of pairing order, the 2D and 3D states begin to have lower energy than the linear 
state. Finally, at $n=0.18$, $p=0.444$ the 3D state emerges as the lowest energy
ground state. 

We were able to perform calculations on moderately sized 3D simulation cells,
up to $15^3$ sites,
using GPUs to dramatically speed up the diagonalization. 
Even with
these speed-ups, our search was somewhat limited by the rapidly increasing computational cost. 
To identify a genuine ground state with 3D structure,
care was taken to ensure that the energy difference between the 2D and 3D states was larger 
than any potential finite-size effect. For the case depicted in the rightmost panel of Fig.~\ref{3D_emergence_wlabels}, 
this energy difference was
${\mathcal O}(10^{-6})$, whereas 
the convergence of the energy of both the 2D and 3D states to the thermodynamic limit was 
${\mathcal O}(10^{-7})$, and the energy tolerance on the self-consistency loop was also ${\mathcal O}(10^{-7})$. 
The convergence to the thermodynamic limit was determined by comparing the energies from
calculations 
using 100 ${\mathbf k}$-points in each direction to calculations using 200 ${\mathbf k}$-points in
each direction for both 2D and 3D structures.

This result demonstrates the existence of
a ground state with LO order of a 3D structure. 
The overall trend suggested by our results is that higher dimensional ground states become 
increasingly stable, at relatively large polarization, with decreasing density near the onset of pairing 
order, and for $n\lesssim 0.18$ the lowest energy ground state is likely to have a pairing order parameter 
with 3D spatial structure.

The 3D structure we observe corresponds to an order parameter that is the sum of six plane waves, 
as described in Sec.~\ref{methods}.  The set of $\qm$ vectors is $|\qm|\{(\pm1,0,0),(0,\pm1,0),(0,0,\pm1)\}$. 
It has been suggested \cite{color_superconductivity} that in the Fermi gas regime the most energetically 
favorable structure is a sum of eight plane waves of the $(1,1,1)$ variety. As discussed in Sec.~\ref{methods}
and indicated by Fig.~\ref{phase_diagrams_horizontal}, we expect solutions with 
$\qm\propto(1,1,1)$ to be stable only at small polarizations. However smaller polarizations result in 
a smaller $|\qm|$, as pictured in Fig.~\ref{U4.0_dens0.96_dens0.60_dens0.24_orderp_q_slice}, and 
a smaller $|\qm|$ corresponds to a longer wavelength pairing wave. This would require even larger
3D simulation cells, and thus lies outside the parameter region in which we have explored possible
3D structures.
 
\section{Summary}
\label{sec:summary}

We have carried out a systematic study of the phase diagram of spin-imbalanced
fermions
with attractive interactions in a 3D lattice.
The phase space can be divided
into two qualitatively distinct regimes, the optical lattice regime at high density
and the Fermi gas regime at low density. In the optical lattice regime our survey
involves detailed, fully self-consistent HFB calculations in which 
great care is taken to reach the true ground state at thermodynamic limit.
The phase diagram in this regime was determined for 
up to intermediate interaction strengths.
We find that the system favors linear pairing order of the LO type. At $U/t=3$ the pairing vector $\qm$
is directed along $(0,0,1)$, and at $U/t=4$ and $U/t=5$ there is a transition
from states with $\qm$ along $(0,0,1)$ at low polarizations to $\qm$ along  $(1,1,1)$
at intermediate to high polarizations. 
The real and momentum space properties of these phases are determined.
At low polarizations and high to intermediate 
densities the pairing wave is characterized by the presence of domain walls that become
sharper with increasing interaction strength, and the localization of excess spin at the nodes. 
With increasing polarization and decreasing density the pairing wave becomes more sinusoidal 
and the excess spin is less strongly localized. Additionally, pairing becomes more stable
with increasing interaction strength, as evidenced by the growing region of phase
space occupied by ordered phases. 

In the Fermi gas regime we searched for evidence of states with two and three 
dimensional spatial modulation of the order parameter. With the use of GPUs to speed up
the computation, we  performed calculations on simulation cells large enough 
to accommodate both 2D and 3D structures. Our results provide evidence of the emergence of
higher dimensional states, which are most stable for low densities and high polarizations, 
near the onset of pairing order. These states occur as it becomes energetically favorable for the system 
to form pairs with a larger set of pairing vectors. Though our search was limited by the computational 
costs of large cubic simulation cells, our results suggest that for densities below $n\lesssim0.18$ the 
system supports 2D and 3D FFLO states, which makes this an interesting region for future theoretical 
and experimental exploration.  

\subsection*{Acknowledgements}
We acknowledge support from DOE (Grant no.~DE-SC0008627) and NSF (Grant no.~DMR-1409510). 
Computational support was provided by DOE leadership computing through an INCITE grant, and by the 
William and Mary SciClone cluster. We thank Eric Walter for help with computing. 

\bibliography{FFLO3D_references}

\end{document}